\documentclass[pre,twocolumn,superscriptaddress,floatfix,amsmath,a4paper]{revtex4-2}
\usepackage{graphicx}
\usepackage[english]{babel}
\usepackage{amsmath,amssymb}
\usepackage[utf8]{inputenc}
\usepackage{psfrag}
\usepackage{bm}
\usepackage{xcolor}
\usepackage{hyperref}
\usepackage{appendix}
\usepackage[normalem]{ulem}
\usepackage{lipsum}

\begin{document}


\title{Acculturation and the evolution of cooperation in spatial public goods games}


\author{Alessandra F. Lütz}
\email{sandiflutz@gmail.com}
\affiliation{Departamento de Física, Universidade de Minas Gerais, 31270-901, Belo Horizonte MG, Brazil}
\author{Marco A. Amaral}
\affiliation{Instituto de Humanidades, Artes e Ciências, Universidade Federal do Sul da Bahia, 45988-058, Teixeira de Freitas BA, Brazil}
\author{Lucas Wardil}
\affiliation{Departamento de Física, Universidade de Minas Gerais, 31270-901, Belo Horizonte MG, Brazil}
\date{\today}

\begin{abstract}
Cooperation is one of the foundations of human society. Many solutions to cooperation problems have been developed and culturally transmitted across generations. Because immigration can play a role in nourishing or disrupting cooperation in societies, we must understand how the newcomers' culture interacts with the hosting culture.  Here, we investigate the effect of different acculturation settings on the evolution of cooperation in spatial public goods games with the immigration of defectors and efficient cooperators. Here, immigrants may be socially influenced, or not, by the native culture according to four acculturation settings: integration, where immigrants imitate both immigrants and natives; marginalization, where immigrants do not imitate either natives or other immigrants; assimilation, where immigrants only imitate natives; and separation, where immigrants only imitate other immigrants. We found that cooperation is greatly facilitated and reaches a peak for moderate values of the migration rate under any acculturation setting. Most interestingly, we found that the main acculturation factor driving the highest levels of cooperation is that immigrants do not avoid social influence from their fellow immigrants. We also show that integration may not promote the highest level of native cooperation if the benefit of cooperation is low.
\end{abstract}


\maketitle

\section{Introduction}

Humans are cooperators and we need each other to succeed~\cite{nowak:book:2011}. The major transitions in evolution could not have unfolded without the repeated, cooperative integration of lower-level entities into higher-level units~\cite{maynard-smith:book:1995a, szathmary_jtb97, Pennisi2005}. Cooperation shape life as it is, from genetic changes that triggered the transition from unicellular to multi-cellular organisms to cultural changes that mould humans into a global society~\cite{Capraro2018}. Nevertheless, the increasing integration of human society places cooperation as a central challenge~\cite{Curry2020, harari2015}. After all, why one should pay the costs of cooperation if others are willing to provide? 
Cooperation can be analysed from several perspectives. Understanding how and why cooperation succeeds is of paramount importance. Regarding the ``why'' questions, we are looking at ultimate explanations for the evolution of cooperation and Evolutionary Game Theory  is a powerful mathematical framework to provide such answers~\cite{Perc2017}. By modelling strategies from classical game theory in an evolutionary context, where more adapted strategies are able to spread through the population, this theoretical framework is able to show how and why different strategies can propagate in diverse scenarios~\cite{Nowak2006}.

In Evolutionary Game Theory, cooperation is often modelled as a strategy in a Public Goods Game (PGG)~\cite{szolnoki_epl10}. This game offers the possibility to analyse the main dilemma of cooperation: the tension between self-interest and the social optimum. In the PGG, individuals interact in a multi-player game with two strategies: cooperation or defection. All cooperators contribute with a given amount to a common pool of resources. The total contribution is multiplied by a  factor $r>1$ (the multiplicative factor represents the synergistic effect of the collaboration) and then it is equally divided among all individuals (including defectors).
If all agents contribute, everyone obtains more than the individual contribution. If no one contributes, no one will get a better payoff. Nevertheless, if a single individual believes that the peers will cooperate and contribute, this individual will have a big incentive to defect, keeping his/her initial endowment while receiving the benefits of the collective pool. If all agents follow this rationale, no one will contribute and the whole population ends in the so-called ``tragedy of the commons''~\cite{hardin:Science:1968, killingback_prsb06}.

The PGG shows that it is challenging to establish cooperation if competition is at stake because defectors are always better off than cooperators. However, if mechanisms like spatial structure are present, cooperators may have a chance depending on the cost to benefit ratio of participating in the PGG~\cite{nowak:Nature:1992b, szabo:PRL:2002, wakano:JTB:2007, santos:Nature:2008, killingback_prsb06}. Besides spatial structure, that gives rise to network reciprocity, other major mechanisms that help cooperation are the punishment of defectors, the rewarding of cooperators and the ability to not participate in the game~\cite{Flores2020, Krawczyk2018, Javarone2016d, Wardil2017, Fang2019, Perc2016, Szabo2007, perc_bs10, perc_jrsi13, Wang2012, vainstein_pa14, Zhao2020}. 
In spatial PGG, cooperation can thrive if the multiplicative factor is high enough. This multiplicative factor represents how a private resource is converted into a public good. For example, more efficient behaviours can improve recycling without extra costs;  more efficient government systems can avoid unnecessary bureaucracy, leading to more efficient use of taxes for the population, etc. 
Notice that cooperation never thrives in the absence of spatial structure and other mechanisms like punishment or reward. Thus, one trivial way to promote cooperation in society would be to increase the multiplicative factor. However, this may not be easy due to the deep cultural roots of social behaviours.

In human societies, immigration plays a central role both in economics and in how the individuals interact with one another~\cite{migration_climate, immigration-selection, immigration-economicgrowth}.
The effect of human migration was analysed in~\cite{Berry:1997uv, berry2006immigrant,richerson2008migration,boyd2009voting,boyd1988culture}. In~\cite{richerson2008migration}, the authors discuss the role of migration in cultural evolution. In particular, they show how such changes tend to have a more durable and positive impact when there is some integration between the immigrant and the native cultures.

Returning to the problem of cooperation, could one society benefit from the arrival of super-cooperators from cultures where pro-social behaviour is highly valued, so that the local production of public goods is improved? How two different cultures should interact so that the new culture can improve native cooperation?
The first question of whether super-cooperation can benefit native cooperation was investigated in~\cite{LuAmWar21}. In this work, efficient cooperators are defined as players whose contribution to the public goods is multiplied by $\alpha r$, with $\alpha>1$. The native population is composed of defectors and standard cooperators. It was shown that the arrival of super-cooperators and defectors immigrants improve native cooperation in settings where cooperation would not be sustained in the absence of migration. Moreover, it was shown that native cooperation reaches a peak for moderate values of migration rate. In~\cite{LuAmWar21}, it was assumed that the immigrants are perfectly integrated into the new society and the immigrants are fully opened to the social influence from the native society and the fellow immigrants. This brings us to the second question: how do the immigrants respond to the hosting culture? 

The arrival to a new country is not an easy process. The cultural shock, linguistic barriers, among other factors, are part of a phenomenon called acculturation. Acculturation refers to changes that occur as a result of continuous first-hand contact between individuals of differing cultural origins~\cite{Ward:2001ve}. For example, social norms can evolve in society through imitation processes akin to how evolutionary game theory dynamics works~\cite{boyd2002group}. Such cultural norms can be beneficial to a given society and can be seen as a form of acculturation.
Researches have been conducted to better understand the main aspects and consequences of the acculturation process in terms of both psychological and socio-cultural adaptation~\cite{berry2006immigrant,ward2008thinking}. Psychological adaptation is related to aspects as anxiety and depression level, self-identity perception, etc, while socio-cultural aspects refer to the immigrants' behaviour in socio-cultural activities. In~\cite{berry2006immigrant}, immigrant adolescents were interviewed. According to their answers, they were categorized into four groups, based both on their psychological and socio-cultural adaptation, and also their inclinations towards the host country culture and their own: national group, for those that prefer engaging in the host country socio-cultural activities and distance themselves from their cultural heritage; the ethnic, for those who prefer focusing on their ethnic group; integrated, where people want to engage in both their ethnic and also on national socio-cultural activities; and the diffuse group, which is characterized by low attachment for both their ethnic and the host country culture, lack of self-identity and some contradictory answers about their views on acculturation mechanisms. Interestingly, both the integration and the ethnic groups show the best results for psychological and socio-cultural adaptation, the integration mechanism being the best one. 

There are three broad theoretical approaches to the acculturation process:  social identity,  cultural learning and the stress-and-coping approaches~\cite{Ward:2001ve}. In the social identity approach, the individuals identify themselves as members of groups. Thus, questions regarding the worth of maintaining the cultural heritage, or not, are relevant to define acculturation types. In~\cite{Berry:1997uv} the author proposes four acculturation mechanisms, from the point of view of the immigrants. If the immigrants want to maintain their culture while incorporating features of the new one, the acculturation process is called \textit{integration}. If the immigrants leave their homing culture to become progressively identified with the host culture, the process is called \textit{assimilation}. If the immigrants refuse to assimilate the host culture and try to maintain the home culture we have \textit{separation}. Finally, if the immigrants find themselves in a position where they do not want or cannot adopt either the home or the host cultures, the process is called \textit{marginalization}.

In this work, we investigate the effect of acculturation in the evolution of cooperation in spatial public goods games with migration under different acculturation settings. We analyse a population structured on a square lattice subject to random birth and death so that vacant sites allow migration. Natives can choose between cooperation or defection. The immigrants can also choose the efficient cooperation strategy. The strategy frequency evolves through imitation. The natives can imitate everyone. However, who the immigrants imitate depends on the acculturation mechanism. Certainly, we could extend the model to analyze acculturation from the perspective of the native culture, but the point here is the unidirectional influence coming from a foreign culture.

Here, we show that cooperation is greatly facilitated and reaches a peak for moderate values of migration rates under all four acculturation settings. The highest native cooperation value is achieved under the integration and the separation settings, which are mechanisms featuring the possibility for immigrants to imitate other efficient cooperators. If immigrants are not able to imitate other immigrants, then cooperation is lower. We also see that the possibility of immigrants to imitate, or not, the natives introduces subtle differences in the population dynamics, for example,  it can cause inversions of which setting is the best at promoting native cooperation. We present the results in the following order. First, we briefly discuss some results to compare the four acculturation settings in a more generic way. Then, we analyse in depth the role of being able to imitate other immigrants and the role of being able to imitate natives. Finally, we provide some results on the impact of the acculturation settings on the social welfare. 

\section{The model}

We analyse cooperation in the context of spatial public goods games with three strategies: defection ($d$), cooperation ($c$), and efficient cooperation ($e$). In a public goods game, a cooperator pays a cost $c$ to produce a public good (we use $c=1$ without loss of generality). The contribution of each cooperator in the group is multiplied by a factor $r$, with $r>1$, and the contribution of each efficient cooperator is multiplied by $\alpha r$, where $\alpha$ represents the efficient cooperator capacity to generate more common goods with the same amount of investment ($\alpha>1$). The total contribution is then divided evenly among all participants. Defectors are those that enjoy the benefits without paying the costs. In a group of size $n$, if $n_c$ and $n_{e}$ are, respectively, the number of standard and efficient cooperators, the payoffs are given by
\begin{eqnarray}
\pi_c=\pi_{e}&=&\frac{rn_c+\alpha r n_e}{n}-1, \label{eq.payoffC}\\
\pi_d&=&\frac{rn_c+\alpha r n_e}{n}, \label{eq.payoffD}
\end{eqnarray}

The population is structured on a square lattice with periodic boundary conditions. Each site is the centre of a public goods game that is played by the individual in the central site and its four nearest neighbours. Thus, each individual plays five public goods games. The total payoff of player $i$, $\Pi_i$, is obtained by summing the payoffs in the five games that player $i$ can participate. 

Regarding the origin of birth, there are two types of individuals: the natives and the immigrants. The natives are individuals born in the population and, the immigrants are those who come from outside. The natives can adopt only defection or standard cooperation. Efficient cooperation is exclusive to immigrants. The assumption is that the efficiency is nurtured only in the original immigrant culture and cannot be transmitted far from the original influences. Both natives and immigrants can adopt defection. In summary, the strategy space available for the natives is $\{c,d\}$ and for the immigrants is $\{e,c,d\}$.

\begin{figure}[h]
\includegraphics[width=\linewidth]{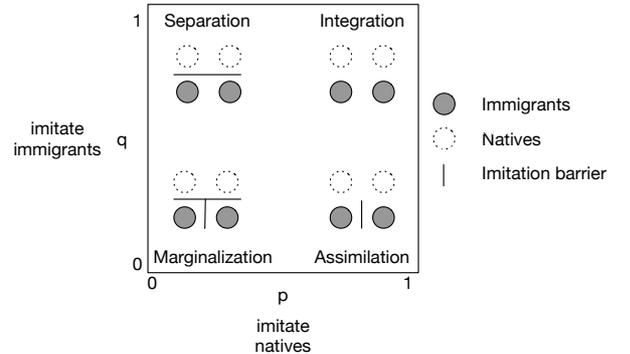}
\caption{Acculturation settings diagram. The probability for an immigrant to imitate a native strategy is $p$ and to imitate another immigrant is $q$. We restrict our analysis to the four vertexes of the $(p,q)$ diagram, namely: Integration $(p=q=1)$, Separation $(p=0,q=1)$, Assimilation $(p=1,q=0)$, and Marginalization $(p=q=0)$. In the diagram, the dotted circles represent native agents and the dark grey circles represent immigrants. The black lines represent the impossibility of imitation between the agents for a given acculturation setup. Natives can imitate any player.}
\label{culture_diagram}
\end{figure}

The evolution of cooperation is determined by the imitation rule, which is a bounded rationality model where individuals try to access the best strategy by comparing their payoffs to the payoffs of other players. Such a process leads strategies that yield higher payoffs to spread at higher rates~\cite{xu_z_pre09}. However, depending on how one individual manages to adapt to the new context after migration, there can be four scenarios of acculturation.
In the \textit{integration} setting, the immigrants imitate both natives and other immigrants. In the \textit{assimilation} setting, because the immigrants cannot maintain their own identity, they imitate only the natives. If immigrants want to preserve their own culture, they use the \textit{separation} setting. Finally, if there is little interest in maintaining the original culture or assimilating the local one, the setting is called \textit{marginalization}. 
These four settings can be mapped to a more general setup where immigrants imitate natives with probability $p$ and other immigrants with probability $q$. The pair $(p,q)$ defines the acculturation setting space. In our work, we analyze populations where only one mechanism is present through the population and we restrict our analysis to the four vertices of the $(p,q)$-space: $(1,1)$, $(1,0)$, $(0,1)$, and $(0,0)$. Figure~\ref{culture_diagram} illustrates the four acculturation settings. Although we could also analyze how the native culture adapts to the arrival of immigrants, the focus here is only on the acculturation settings on the immigrant side.

For the demographic part of the model, we considered random birth and death events so that empty spaces are present and immigration is possible. Individuals reproduce with a probability $\beta\rho_v$, where $\beta$ is a system parameter controlling the reproduction and $\rho_v$ is the fraction, $N_v/N$, of vacant sites in the system due to the deaths. The offspring inherits their parent’s strategy, except that the strategy of the efficient cooperators' offspring is standard cooperation. This is so because of the cultural integration they are subjected to. We must mention that there can be different mechanisms of offspring integration, but here we considered the simplest scenario. The parameter $\gamma$ controls the population density. Individuals die with a probability $\gamma \beta (1-\rho_v)$. In this way, the average population density fluctuates around $(1-\gamma)$ for the immigration rates used in our simulations. Notice that, because players can have empty sites around them, the size of public goods groups can be smaller than five. Empty sites receive immigrants with a probability $\mu\rho_v/(1-\rho_v)$. Thus, at each time step, approximately $\mu N_v$ immigrants arrive, $\mu$ being a system parameter that controls the immigrants influx. Notice that the immigration flux depends not only on the migration coefficient $\mu$ but also on the global density. The constant influx of immigrants is comprised of half defectors and half efficient cooperators. See the appendix for the detailed implementation of the Monte Carlos steps and further details on all the parameters used.
In the following analysis, we  distinguish five types of individuals: native cooperators ($c_0$), native defectors ($d_0$), immigrant cooperators ($c_i$), immigrant defectors ($d_i$), and immigrant efficient cooperators ($e$).

\section{Results}

\begin{figure}[t]
\includegraphics[width=8cm]{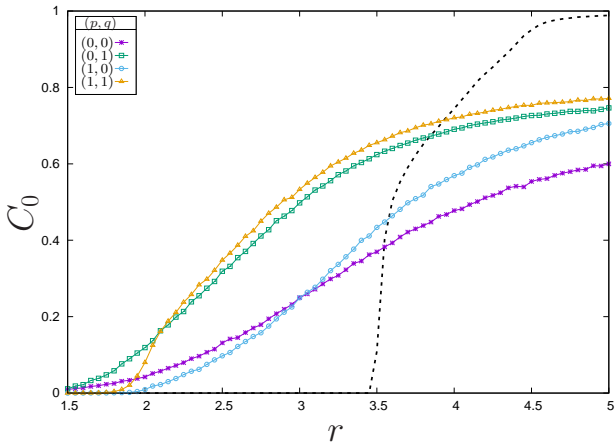}
\caption{Asymptotic density for the native cooperators as a function of the factor $r$, considering a migration coefficient of $\mu=0.05$. The dashed black line corresponds to the native cooperator density when there is no immigration, $\mu=0$, while each of the other curves is associated with a system where immigrants use a different acculturation setting, $(p,q)$, with $p,q\in\{0,1\}$.}
\label{fig.c0Xr}
\end{figure}

In our model, cooperation can be maintained in the absence of migration only if $r\leq3.4$, as shown in Fig.~\ref{fig.c0Xr}. 
However, if migration is allowed, native cooperation can be maintained for lower values of $r$ under all acculturation settings. 
Notice that we are measuring the fraction of native individuals adopting cooperation, not the total fraction of cooperation. Any growth in the overall cooperation caused by an increase in the proportion of immigrant efficient cooperators is a trivial phenomenon. 

For $r\leq3.4$, any type of acculturation setting is beneficial for native cooperation, even the marginalization one. This is so because in our model natives can always imitate immigrants, regardless of the acculturation setting. Thus, the possibility of imitating immigrants paves the way for the positive influence of super-cooperators.

\begin{figure*}
\includegraphics[width=15cm]{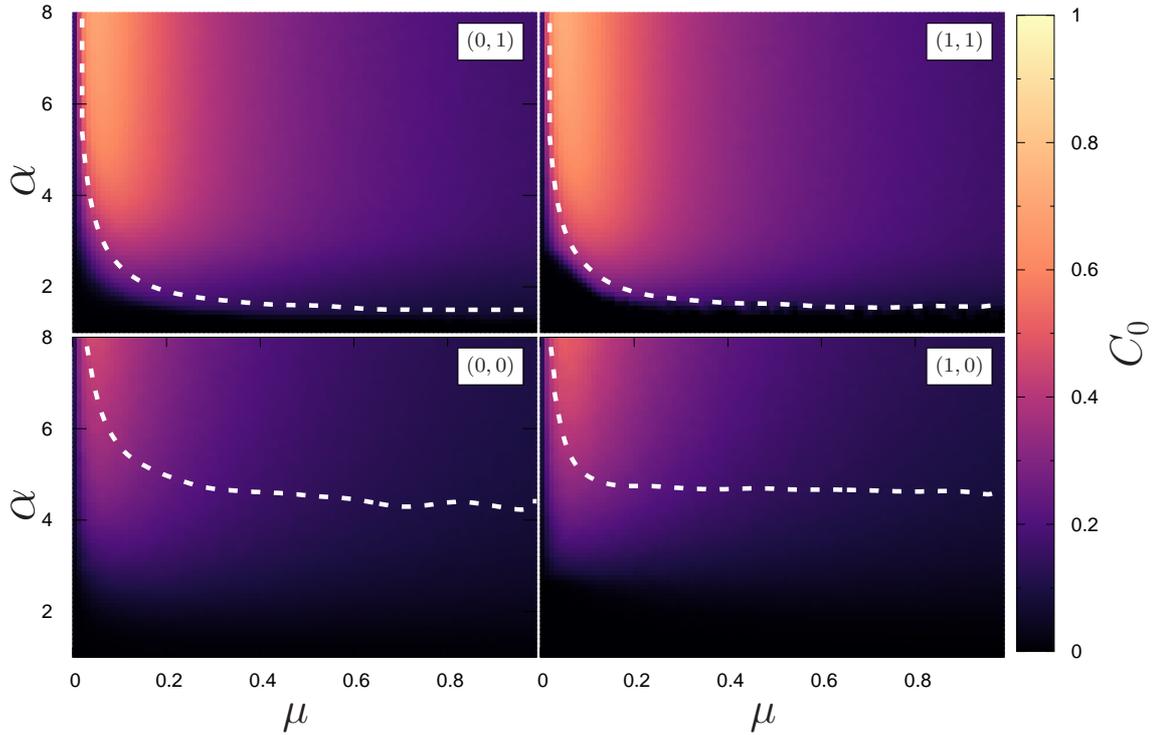}
\caption{Diagram of the native cooperation density as a function of the migration coefficient, $\mu$, and the $\alpha$, for all the acculturation settings. Here we considered $r=3$. The dashed white line delimits the region above which cooperators are the majority of the population, while the color scale indicates the final native cooperation level. On both of the top row diagrams, where $q=1$, the area above the dashed line is larger than those on the bottom row, where $q=0$. Additionally, comparing the colors on the diagrams, it is clear that there are more native cooperators on the top row diagrams, for almost all the parameters than on the bottom ones. Thus, the settings where immigrants can imitate other immigrants, $q=1$, are better not only for cooperation in general but also for native cooperation.}
\label{fig.diag}
\end{figure*}
In Fig.~\ref{fig.diag}, we show a diagram for the density of native cooperators as a function of the migration coefficient, $\mu$, and the immigrants' efficiency, $\alpha$, for all the acculturation settings. By doing so, we can further analyze the effects of the acculturation setting for different levels of both the efficiency and the number of immigrants in the system. 
The first evident effect is that acculturation settings with $q=1$ (top rows) have larger regions where most individuals are cooperators. Thus, the main driver of any type of cooperation among different acculturation settings is the ability of defector immigrants to imitate efficient cooperators.
On the other hand, the ability of immigrants to imitate native individuals (determined by $p$) has a minor effect. Comparing the first column ($p=0$) with the second one ($p=1$), it is possible to see some subtle differences. In the separation setting ($p=0$, $q=1$), for example, the dark area is smaller for the integration setting ($p=q=1$). The same happen for marginalization ($p=q=0$) in comparison with the assimilation setting ($p=1$, $q=0$). Thus, for $p=0$, the minimum efficiency level necessary for the survival of native cooperators is slightly lower than otherwise. Next, we further explore the specific effects of the parameters $p$ and $q$.

\subsection{The role of $q$}

As mentioned before, for most values of $r$, both the separation ($p=0,q=1$) and the integration ($p=q=1$) settings benefit native cooperation more than the other two settings, where immigrants do not copy other immigrants ($q=0$). In human society, the parameter $q$ can be related to how well immigrants are able to maintain their social ties. The importance of imitating other immigrants lies in the fact that the immigrants that are defectors can become efficient cooperators. If $q=0$, immigrants never copy other immigrants, and so, immigrant defectors can never become efficient cooperators and boost native cooperation. Notice that $q\ne 0$ also allows the reverse effect: efficient cooperators becoming defectors. However, the spatial structure and the higher payoff produced by efficient immigrants give to efficient cooperators the advantage of being the ones that are copied for a broader range of parameters. 

Although, in general, both the integration and the separation settings benefit native cooperation better than the other acculturation settings, that is not the case if the environment gets too harsh for cooperation, that is, for small values of $r$. For example, when $r<2.01$ the acculturation setting that most benefits native cooperators is separation ($p=0, q=1$), while for integration, native cooperators nearly disappear, as can be seen in Fig.~\ref{fig.c0Xr}. Additionally, for $r<3$, marginalization is more beneficial for native cooperation than assimilation, while for $r>3$, assimilation starts to be more beneficial. Thus, the acculturation settings cannot be ranked based only on their benefits for native cooperation without taking into account other crucial characteristics of the system, such as $r$.

\begin{figure*}

\hspace{0.5cm}$(a)$\hspace{7cm}$(b)$

\includegraphics[width=8cm]{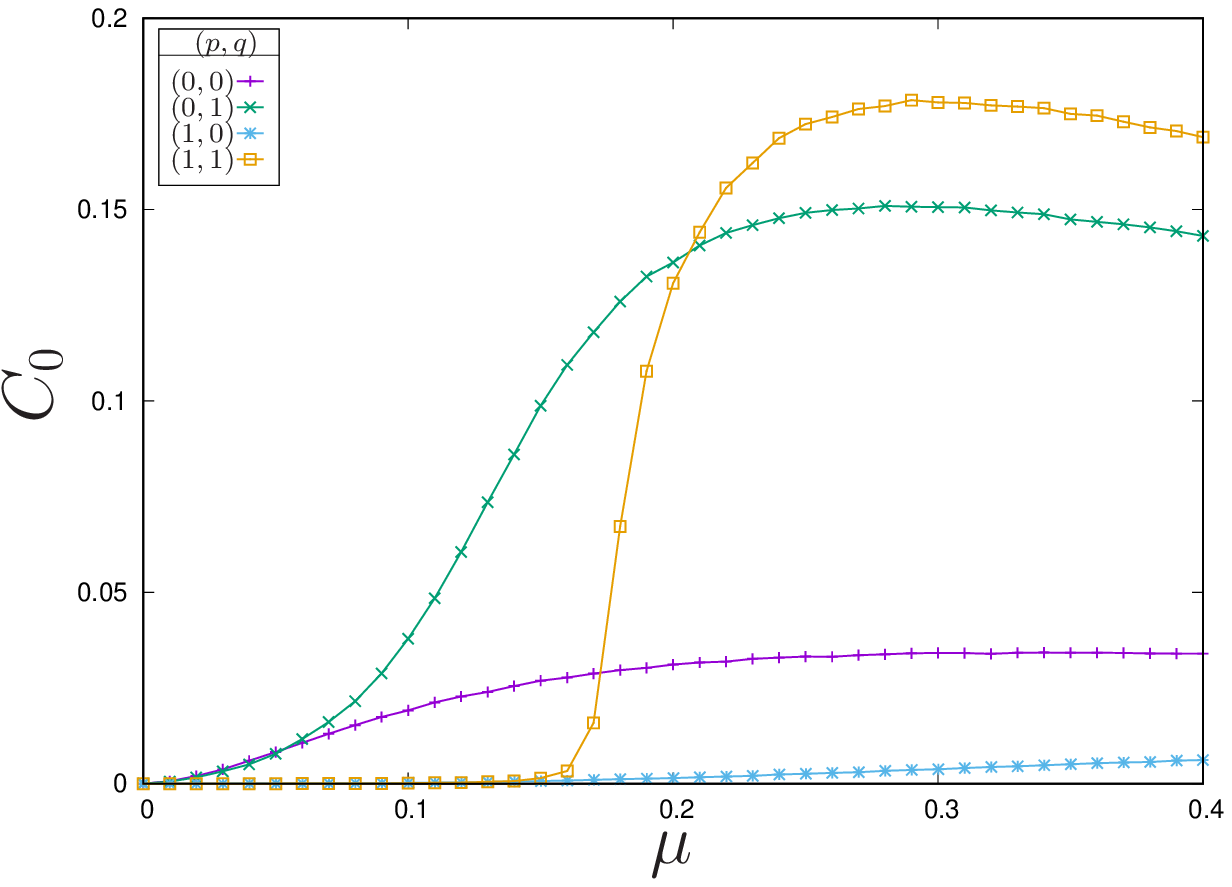}
\includegraphics[width=8cm]{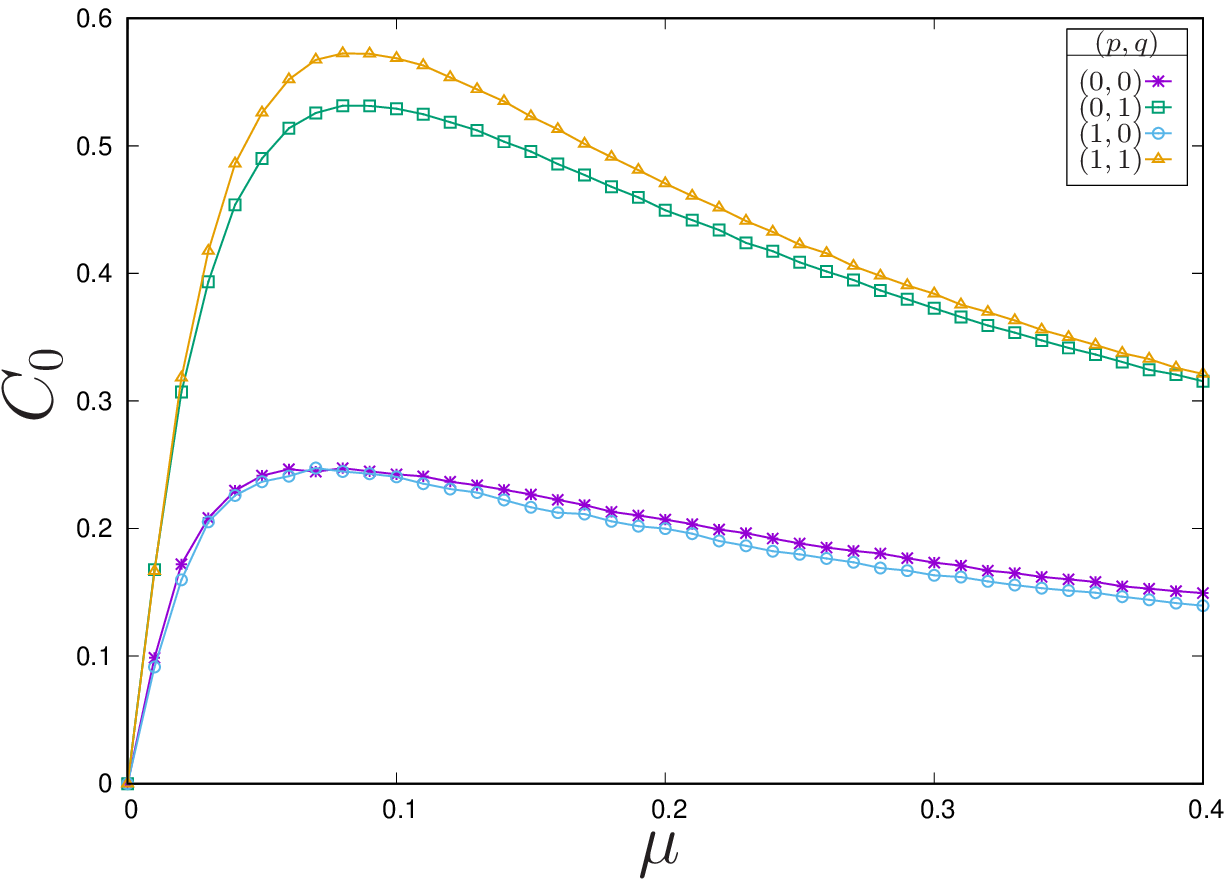}
\caption{Native cooperation density as a function of the migration coefficient, $\mu$, for $a)$ $r=1.5$ and $b)$ $r=3$.
Each curve is related to a different acculturation setting $(p,q)$. For $r=1.5$, the separation ($p=0, q=1$) is the acculturation setting that best benefit native cooperation when $\mu<2.06$. After that, integration ($p=q=1$) starts to be the better one. Such inversion is observed only for small values of $r$, where defectors thrive.}
\label{fig.c0mu}
\end{figure*}

The effect of the migration coefficient, $\mu$, is presented in Figs.~\ref{fig.c0mu}(a) and~\ref{fig.c0mu}(b), for two different values of $r$. In both cases, the optimum positive effect of the increase in the migration coefficient over native cooperation is reached for moderate values of $\mu$, under all acculturation settings. This is so because, for higher values of $\mu$, immigrants become too frequent and compete for space with natives. Such competition dampens the native cooperation growth, even reversing it as the immigration coefficient increases further. In other words, there is an optimal value of $\mu$ to promote native cooperation, irrespective of the acculturation setting. Figure~\ref{fig.c0mu} also shows that the possibility of immigrants to imitate other immigrants, that is, to keep their social ties, has a strong impact on the promotion of higher levels of native cooperation. Notice that if immigrants avoid social influence from their fellows, as in the marginalization and the assimilation settings, the peak of native cooperation decreases. 

Interestingly, the integration setting does not promote the highest level of native cooperators in harsh environments for cooperation, e.g. for low values of the coefficient $\mu$ or the factor $r$ (see Fig.~\ref{fig.c0mu}(a), where $r=1.5$). However, as $\mu$ increases, integration suddenly becomes the best setting for boosting native cooperation, as can be seen for $\mu>0.15$ in Fig.~\ref{fig.c0mu}(a). The reason for such abrupt change can be summarized by the common saying: ``\textit{there is strength in numbers}''. Since the environment is harsh for cooperation, efficient cooperators have to rely only on other fellow efficient immigrants in order to sustain cooperation. When this is not possible, they fall into temptation and become defectors. Notice that this may happen for all the acculturation settings where $p$ or $q$ are nonzero. When $p=q=1$, however, efficient immigrants can imitate both native and immigrant defectors. 
As more efficient cooperators arrive in the system, more frequently they become neighbors among themselves. When this happens, their payoff increases, along with their influence over defectors. In such a scenario, integration is the best option for boosting cooperation since it allows efficient cooperators to influence both native and immigrant defectors.

In the separation setting, the effect of the immigration coefficient is smoother. Because immigrant defectors can only imitate immigrant cooperators, any $d_i$ close to a native cooperator cannot imitate him/her. More than that, such defector may even stay in the way of that native cooperator's influence over native defectors in the neighborhood. Despite that, immigrant defectors that imitate immigrant cooperators, become efficient cooperators. This increases the number of efficient immigrants in the system, strengthening even further their influence. Such a phenomenon cannot happen either in the assimilation or in the marginalization settings. The positive effect of increasing $\mu$ is, therefore, dampened for $q=0$. However, for the marginalization setting, although immigrant defectors can never become cooperators, efficient cooperators are never tempted to become defectors. Thus, when the environment is very harsh for cooperation, e.g. for low values of the coefficient $\mu$ or the factor $r$, separation and marginalization are the best options for boosting native cooperation.

\begin{figure*}[t]

\includegraphics[width=7cm]{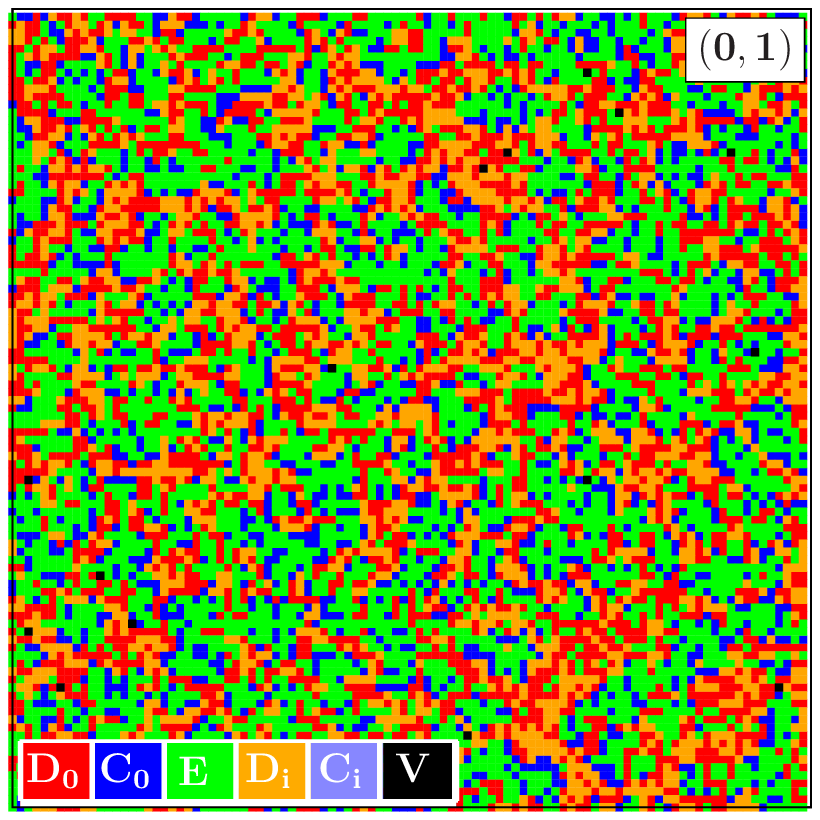}
\includegraphics[width=7cm]{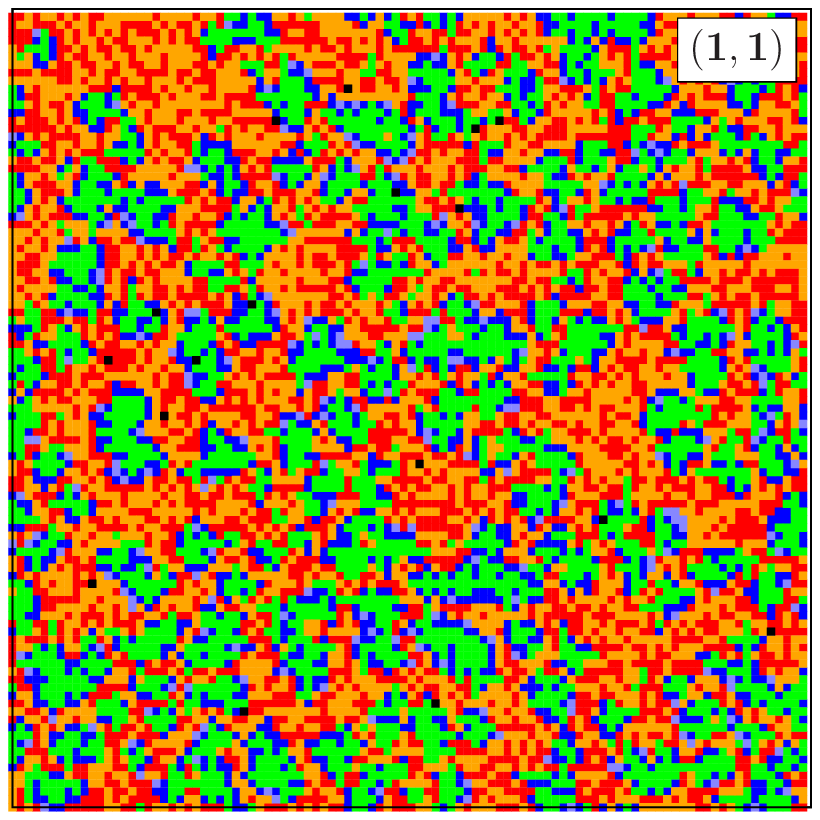}
\includegraphics[width=7cm]{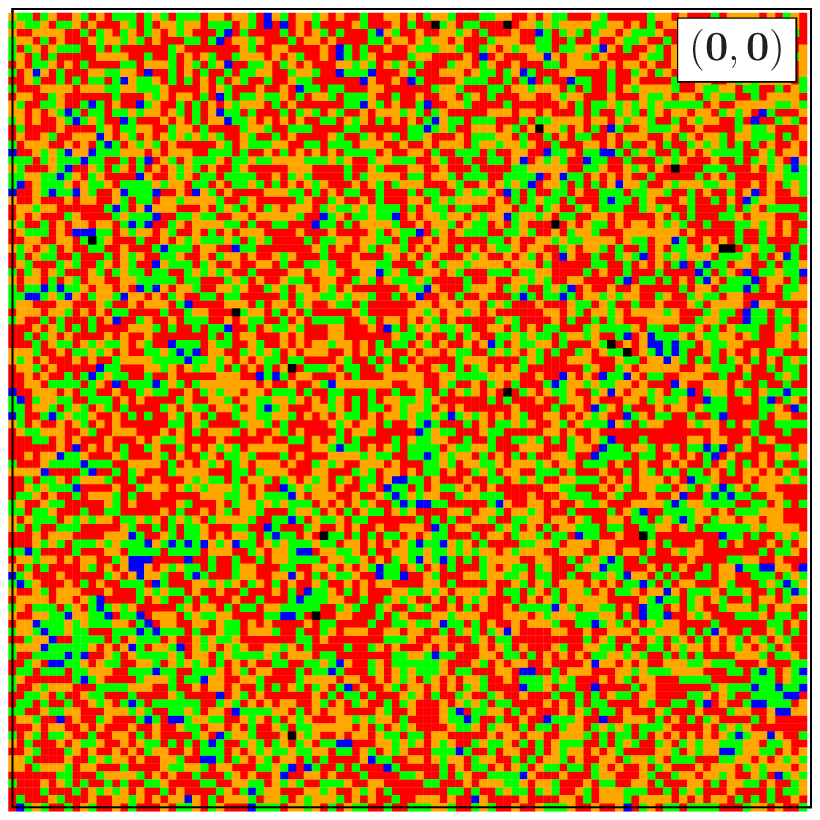}
\includegraphics[width=7cm]{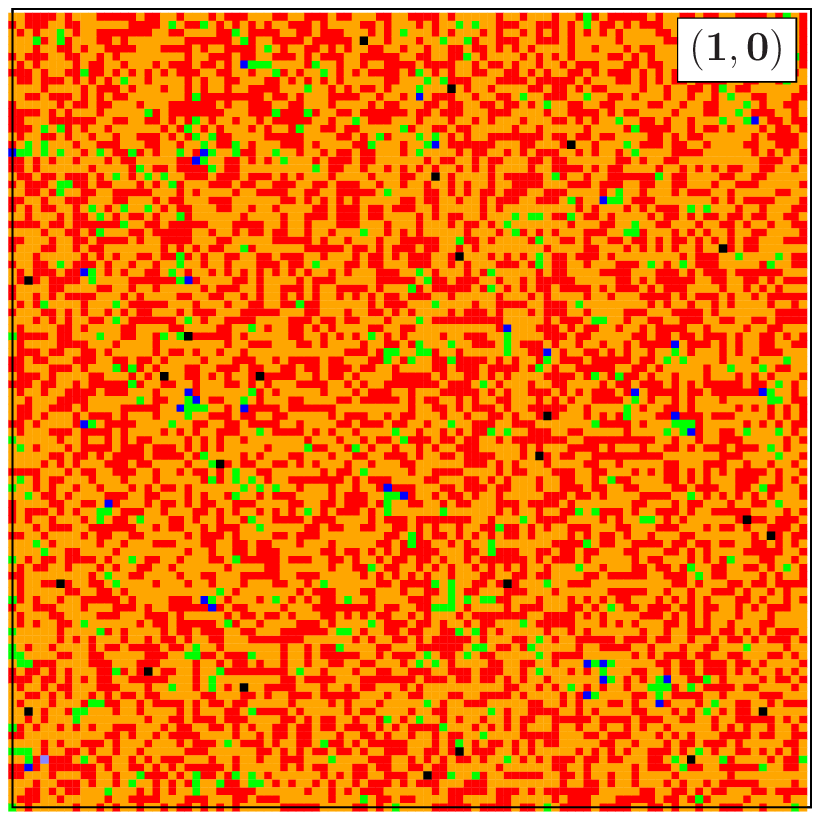}

\caption{Typical spatial configuration for all acculturation settings: separation $(p=0,q=1)$, integration $(p=q=1)$, marginalization $(p=q=0)$ and assimilation $(p=1,q=0)$. The color-coding is as follows: native cooperators (blue), efficient cooperators (green), immigrant regular cooperators (light blue), native defectors (red), immigrant defectors (orange), vacant sites (black). When $q=1$ (top row), immigrants are able to imitate other immigrants, leading to the clusterization of efficient cooperators. This in turn creates a favorable environment for native cooperation to flourish around efficient cooperator clusters. Settings with $q=0$ (bottom row) prohibit immigrants to imitate other immigrants, leading to efficient cooperators to be more disperse and not form clusters. Without efficient cooperative clusters, native cooperators are not able to establish themselves and the cooperation falls drastically. Here we use $r=1.5$ and $\mu=0.25$, but the general effect is present for all studied parameters.}
\label{fig.snapshot}
\end{figure*}

So far, we have seen that the possibility for immigrants to imitate other immigrants is the main factor that drives native cooperation to its highest levels. As shown in Fig.~\ref{fig.c0Xr} and Fig.~\ref{fig.c0mu}, the native cooperation density is higher for the acculturation settings with $q=1$, as long as $r$ or $\mu$ are high enough. Also, the settings with the same value of $q$ yield similar results. 
To better understand the role of $q$ on the strategies' spatial structure, we can compare the system's spatial configurations for all the acculturation settings. Figure~\ref{fig.snapshot} shows a typical snapshot for each of the acculturation settings, considering an environment with low $r$ and high $\mu$. 
%
In the top two snapshots, where $q=1$, it is clear that efficient immigrants (green) form larger clusters than those in the bottom snapshots, where $q=0$. Additionally, the top clusters are usually more heavily surrounded by native cooperators (blue).
The clusters of efficient cooperators are larger for $q=1$ because in these settings the immigrant defectors are able to imitate efficient cooperation and become part of the cluster. 
%
Thus, the ability for immigrants to copy other immigrants, $q=1$, leads to the formation and expansion of efficient cooperation clusters. As those clusters grow, natives that come in contact with them and try to copy their strategy become regular cooperators located at the borders of efficient cooperation clusters, where the payoff is particularly high. In other words, such borders become regions where native cooperation is able to flourish. 

Figure~\ref{fig.snapshot} also shows that when immigrants do not imitate other immigrants, i.e, $q=0$ (bottom snapshots), the efficient cooperators are dispersed in the population. This behaviour remains the same for all the parameters considered.
Without the ability to form clusters, efficient cooperators become surrounded by defectors, both natives and immigrants,  thus being less able to support native cooperation,  which leads to a greater fraction of defectors.
This is the main mechanism behind the sharp differences between the settings with $q=1$ and $q=0$.

\subsection{The role of $p$}

In contrast to the effect of the parameter $q$, the possibility of imitating natives has a minor, but a non negligible, effect on the level of native cooperation: there can be an inversion of the best acculturation setting for boosting native cooperation,  as shown in Figs.~\ref{fig.c0Xr} and~\ref{fig.c0mu}$(a)$. In Fig.~\ref{fig.c0mu}$(a)$, for example, the separation setting ($p=0$, $q=1$) is the best option when $\mu<2.06$, while for $\mu>2.06$, integration ($p=q=1$) is the best one. The assimilation and the marginalization settings may also present and inversion, as seen in Fig.~\ref{fig.c0Xr}.

To understand the mechanism behind such inversions, let us focus on the inversion between integration and separation as $\mu$ increases. For the separation setting, immigrant defectors can only imitate efficient cooperators, never native cooperators. 
Since, for $q=1$ and high enough $\mu$, efficient immigrants form clusters with native cooperators at the borders, the inability of imitating natives ($p=0$) becomes a liability when those conditions are met. 
This is so not only because immigrant defectors can only imitate part of the cooperators in the system, but also because a $d_i$  may stay in the way of a native cooperators' influence over a native defector on the neighborhood. Thus, when $\mu$ is high enough, integration is a better option for boosting native cooperation than separation. When $\mu$ is small, on the other hand, imitating natives is a liability, since most natives are defectors and efficient immigrants are not strong enough to influence them considerably or even resist them. A similar argument holds for the inversion in $r$. 

\begin{figure}
\includegraphics[width=8cm]{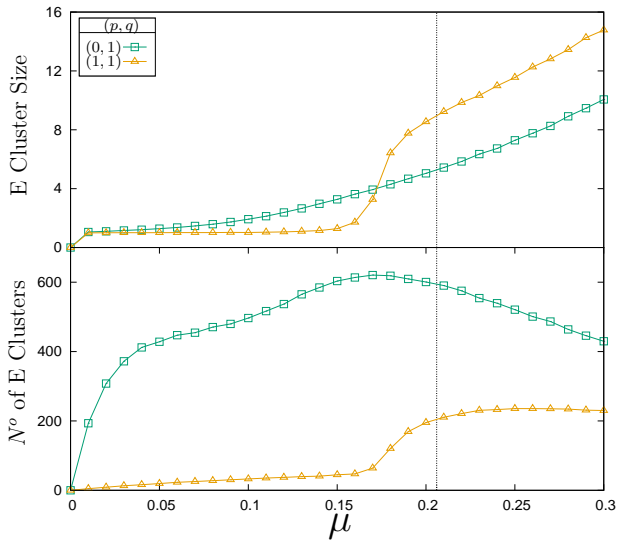}
\caption{ Average cluster size (top) and average number of clusters (bottom) as functions of the migration coefficient $\mu$, for $r=1.5$. The vertical grey line marks the $\mu$ value where the native cooperation inversion occurs. Note that while the cluster size keeps increasing, the average number of clusters starts to diminish after $\mu$ around $0.2$. This indicates that the clusters are so big that they start to merge.}
\label{fig.clustr}
\end{figure}

Because native cooperators grow in the borders of efficient clusters, as shown in Fig.~\ref{fig.snapshot}, one could expect that the inversion is related to the geometry of the clusters: the more fragmented the clusters, the greater the surface where native cooperates can grow. 
Figure~\ref{fig.clustr} compares the average size (top) and the average number (bottom) of efficient clusters for the separation and the integration settings. Both the average size and number of clusters are shown as functions of the migration coefficient. As $\mu$ increases, the average cluster size grows for both acculturation settings, which is expected since more efficient immigrants arrive in the system. 
However, the average number of clusters starts to decrease after a $\mu$ around $0.2$, while the average size keeps increasing. This is a strong indicator that the clusters of efficient cooperators start to merge together after this point. 
As the native cooperators usually grow in the borders of the efficient clusters, such a merging process will leave less space for cooperators to grow, limiting their population. We also see that the native cooperation inversion that happens between the integration and the separation settings occurs around the point where the number of efficient clusters starts to diminish for the separation setting (the inversion point is presented as a vertical grey line in Fig.~\ref{fig.clustr}).

In summary, for low values of the immigration coefficient or the factor $r$, the separation setting ($p=0$ and $q=1$) is more efficient in increasing native cooperation. This is so because for $p=0$ immigrants never imitate natives, which are mostly defectors. However, as $\mu$ increases, cooperation is able to thrive for both the separation and the integration settings. Because there are more native cooperators in the system, the integration setting ($p=q=1$) is better suited to promote native cooperation as immigrant defectors benefit from imitating the cooperative strategy from both natives and immigrants. Additionally, for separation, immigrant defectors close the borders of efficient clusters may stay as defectors for longer periods since they cannot imitate native cooperators. Thus, the effective contact area between native cooperators and defectors may be smaller for the separation setting when compared to integration.

\subsection{The social welfare}

\begin{figure}[t]
\includegraphics[width=8cm]{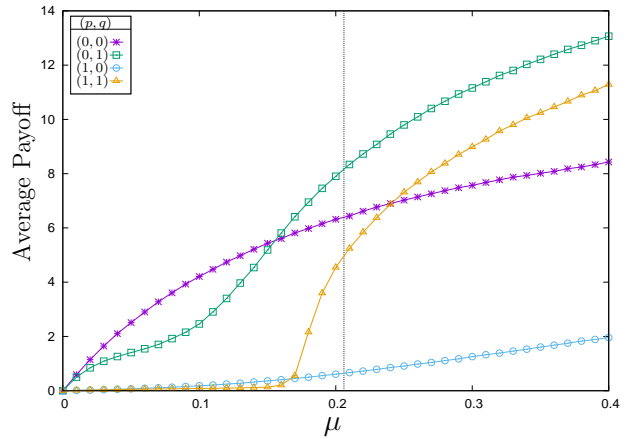}
\caption{Average population payoff (social welfare) as a function of the migration coefficient, $\mu$ for $r=1.5$. Each curve is related to a system with a different acculturation setting $(p,q)$.}
\label{fig.socialwelfare}
\end{figure}

The improvement of the social welfare can be one of the positive consequences of immigration. Thus, let us look at the social welfare under the four types of acculturation. Here, we measure the social welfare as the average payoff in the population. The average payoff as a function of $\mu$ is shown in Fig.~\ref{fig.socialwelfare}. The separation setting generates the highest average payoff for most values of $\mu$, even for those cases ($\mu>0.206$) where the integration setting is the one with the largest population of native cooperators (see Fig.~\ref{fig.c0mu}). This happens because the separation setting is able to maintain a higher fraction of efficient cooperators in the population for all values of $\mu$. In contrast, when $\mu$ is low, the marginalization setting yields the highest social welfare. The reason is that at low immigration rates the overall level of cooperation is small, and the marginalization setting prevents efficient immigrants from imitating defectors. Since cooperators are the only ones that produce public goods, the maintenance of a minimal number of cooperators is necessary to guarantee a basal level of social welfare. Interestingly, because cooperation is mostly adopted by the immigrants, the welfare generated by the public goods is mostly enjoyed in the immigrants' neighborhoods.
Notice that, for large $\mu$, the social welfare for the separation and the integration settings becomes more similar than the other two cases. For higher values of $r$, this similarity is even greater, reinforcing the fact that the immigrant's ability to imitate efficient cooperation is a strong factor in the dynamics.
  
\section{Discussions and conclusion}

The problem of human cooperation is an exciting open question in many disciplines and evolutionary game theory has proven an excellent tool to integrate the different mechanisms that can affect cooperation. The spatial public goods game is a natural framework to investigate how cooperation evolves in the presence of immigration. However, because human cooperation strongly depends on social norms, understanding how the heritage culture interacts with the hosting culture is important to understand the impact of immigrants on cooperation.

In this work, we investigated how different forms of acculturation impacts the evolution of cooperation driven by super-cooperator immigrants. We analyzed four types of acculturation settings:  marginalization (immigrants cannot imitate either natives or immigrants),  assimilation (immigrants imitate only native agents), separation (immigrants imitate only other immigrants) and, integration (immigrants can imitate both natives and immigrants).

We found that the dominant parameter is the possibility of immigrants imitating other immigrants, $q$ in our model. Without this capacity, the overall cooperation level is greatly diminished since efficient cooperators are not able to form cooperation clusters and end up surrounded mostly by defectors. On the other hand, when immigrants are able to imitate other immigrants, compact clusters of efficient cooperators are formed, sustaining native cooperation on its borders and increasing the general cooperation level.

We also observe that the possibility of immigrants imitating natives, the parameter $p$, has a more subtle effect. The inability of imitating natives can actually prevent immigrants from adopting the native's defection strategy in scenarios where it is harsh for cooperation to flourish. In such scenarios, the separation acculturation setting is able to prevent native defection from invading the immigrant population, while at the same time the efficient cooperation strategy is able to spread in both native and immigrant populations. On the other side, when parameters are such that cooperation is favored, imitating natives boosts cooperation as immigrants are allowed to imitate native cooperation.
%
%
%

Certainly, the acculturation process has two sides: how the immigrants adapt to the hosting culture and how the natives respond to the arrival of individuals culturally distinct \cite{Berry:1997uv}. In this work, we investigate only the first question because the goal was to ask whether immigrants coming from cultures where cooperation is very efficient can work as role models to the local population. Another relevant aspect of acculturation is that the maintenance of heritage culture also depends on how culture is transmitted through generations. Here, we considered that the first generation is completely integrated into the hosting society. There is evidence that young and those born within the hosting culture assimilate local culture more quickly than those born overseas \cite{Mendoza:1989uk,Cortes:1994vw}.

The relations between cooperation and migration are very complex and cannot be fully described by a simple mathematical model. Here, our goal was to focus on a single aspect of the phenomenon to better understand how different acculturation settings can affect cooperation. Although policies are hardly based solely on simplified models such as ours, these simple models call attention to potential fields of intervention. Moreover, other aspects should be studied if one wants to have a better picture of the many dimensions involved in the problem. Some examples are wealth distribution, inequality or, how wealthier individuals have different behaviors in competitive games~\cite{Shuler2019}.

In summary, immigrants coming from societies with strong pro-social norms may act as positive role models to boost native cooperation. Even if defectors may also come together, the impact of the super-cooperators is much greater. Care must be taken so that the valuable cultural assets from the new culture are properly accommodated to the hosting culture. By doing so, the best of the two cultures adds up and everyone can enjoy the benefits of cooperation. Particular attention should be paid to how immigrants interact among themselves so that the cooperation culture coming with them is not undermined.

\begin{acknowledgments}
This research was supported by the Brazilian Research Agency CNPq (proc. 428653/2018-9), the Brazilian Research Agency CAPES (proc. 88882.463226/2019-01) and the Minas Gerais State Agency for Research and Development FAPEMIG.
\end{acknowledgments}

\section*{Contributions}
All authors contributed equally to the paper.

\bigskip

\textbf{Data Availability Statement:} This manuscript has no
associated data or the data will not be deposited. 
\appendix

\section{Computer simulations method}
The simulations are run in a square lattice of size $N=100\times100$ starting fully occupied with a fraction of $0.2$ defectors and $0.8$ standard cooperators. Note that the initial state does not affect the equilibrium states of the system. In each Monte Carlo step (MCS), the following steps are repeated $N_o$ times, with $N_o$ being the number of occupied sites. First, a player $i$ is randomly chosen to imitate an individual $j$ in the neighborhood with a probability
\begin{equation}
p_{i \rightarrow j}=max \left\{  \frac{\Pi_{j}-\Pi_{i}}{\Delta \Pi_{max}},0 \right\} 
\end{equation}
where $\Delta \Pi_{max}$ is the maximum payoff difference considering all possible combinations of allowed payoffs, which is included to normalize the probabilities~\cite{Szabo2007}. Here, $\Pi_i$ is the total payoff of player $i$, obtained by summing the payoff from all games that player $i$ participates. Notice that the probability $p_{i \rightarrow j}$ does not take into account irrationality. The Fermi imitation rule~\cite{Szabo2007}, on the other hand, is more realistic and considers irrationality. The conclusions of our model remain the same if we use the Fermi imitation rule.

After the imitation step, another individual and one of its first-neighbor sites are randomly chosen. If the neighbor site is empty, the individual reproduces with probability $\beta$. Third, another individual is chosen randomly and dies with probability $\gamma \beta$. Last, for the immigration step, a site is chosen randomly and, if it is empty, it receives an immigrant with probability $min\{1,\mu/\rho_o\}$, where $\rho_o$ is the fraction of occupied sites and $\mu$ is the immigration coefficient.  As long as $\mu \le \rho_o$, the factor $\rho_o$ guarantees that at each MCS the number of incoming immigrants is, on average, equal to $(\mu/\rho_o)(1-\rho_o)N_o=\mu(1-\rho_o)N=\mu N_v$, where $N_v$ is the number of vacant sites. 

We consider a transient time around $10^4$ to $10^5$ MCS, after which we average the measures over the last $1000$ steps. The results are further averaged over $100$ independent samples.
In the current paper, we fixed $\alpha=4, \gamma=0.005, \beta=0.2$, unless stated otherwise. A further investigation on the effects of varying these parameters in a scenario without acculturation can be found in~\cite{LuAmWar21}.

\bibliographystyle{ieeetr}

\end{document}